\documentclass[11pt]{article}
\usepackage{a4,bm,epsfig,booktabs}
\usepackage{setspace}
\usepackage[T1]{fontenc}

\newcommand{\Wegdamit}[1]{}
\newcommand{\AP}{{\mathrm{AP}}}

\usepackage{ifthen,array}
\newcommand{\ams}{\usepackage{amsfonts,amssymb,amsmath}}

\ams
\allowdisplaybreaks[3]
\newlength{\textwidthorig}
\newlength{\oddsidemarginorig}
\newlength{\textheightorig}
\newlength{\topmarginorig}
\def\seitenlaengenabsolut#1 #2 #3 #4 {\setlength{\textwidth}{#1}
                                      \setlength{\oddsidemargin}{#2}
                                      \setlength{\textheight}{#3}
                                      \setlength{\topmargin}{#4}}
\def\seitenlaengenrelzustandard#1 #2 #3 #4 {\setlength{\textwidth}{\textwidthorig+#1}
                                            \setlength{\oddsidemargin}{\oddsidemarginorig+#2}
                                            \setlength{\textheight}{\textheightorig+#3}
                                            \setlength{\topmargin}{\topmarginorig+#4}}
\def\seitenlaengenrelzuvorher#1 #2 #3 #4 {\addtolength{\textwidth}{#1}
                                          \addtolength{\oddsidemargin}{#2}
                                          \addtolength{\textheight}{#3}
                                          \addtolength{\topmargin}{#4}}
\newcommand{\standardseite}{\seitenlaengenrelzuvorher2.2cm -0.8cm 1.8cm -1.5cm }   %
\standardseite

\newlength{\laengespatium}

\newcommand{\nach}{\longrightarrow}      

\newcommand{\auf}{\longmapsto}           
\newcommand{\txtauf}[1]{\auf}            
\newcommand{\impliz}{\Longrightarrow}    
 
\newcommand{\invimpliz}{\Longleftarrow}  
\newcommand{\gegen}{\rightarrow}         
\newcommand{\ident}{\equiv}              
\newcommand{\teilmenge}{\subseteq}       
\newcommand{\obermenge}{\supseteq}       
\newcommand{\aeqrel}{\sim}               

\newcommand{\einschr}[1]{{}\arrowvert_{#1}}      
\newcommand{\dirsum}{\oplus}           

\newcommand{\betraganpass}[1]%
           {\left| #1 \right|}           
\newcommand{\bigbetrag}[1]%
           {\bigl|{#1}\bigr|}            
\newcommand{\betrag}[1]%
           {|{#1}|}                      
\newcommand{\betragnichtanpass}[1]%
           {\mid #1 \mid}                
\newcommand{\norm}[1]%
           {{}{\parallel}#1{\parallel}{}}      
\newcommand{\erww}[1]%
           {\langle #1 \rangle}          
\newcommand{\skalprod}[2]%
           {\langle #1,#2 \rangle}       
\newcommand{\supnorm}[1]{{\norm{#1}_\infty}}        
\newcommand{\quer}{\overline}            
\newcommand{\dach}{\widehat}             
\newcommand{\inv}[1]{\frac{1}{#1}}       
\newcommand{\re}{\text{Re }}                           
\newcommand{\del}{\partial}                            
\newcommand{\I}{\text{i}}                              
\newcommand{\EINS}{{\boldsymbol{1}}}                   
\newcommand{\field}[1]{\mathbb{#1}}                    
\newcommand{\C}{{\field{C}}}                           
\newcommand{\N}{{\field{N}}}                           
\newcommand{\R}{{\field{R}}}                           
\newcommand{\boundfkt}{{\cal B}}                       
\newcommand{\rnkl}[2]{\raisebox{-0.4ex}{$#1$}%
\raisebox{-0.12ex}{{\large$\setminus$}}\,#2}   
\newcommand{\agb}{{\overline{{\cal A}/{\cal G}}}}      
\newcommand{\agbfact}[1][]{\text{$\agb/\!\aeqrel$}}    
\newcommand{\Gb}{{\overline{{\cal G}}}}                

\newcommand{\qa}{{\quer{A}}}                           

\newcommand{\holgr}{{\mathbf H}}                       
\newcommand{\bz}{{\mathbf B}}                          

\newcommand{\LG}{{\mathbf{G}}}                         
\newcommand{\aeqrelzush}[1][]{\sim}                    
\newcommand{\alg}{\mathfrak{A}}                          
\newcommand{\blg}{\mathfrak{B}}                          
\newcommand{\malg}{{\cal M}(\alg)}                     

\newcommand{\nklza}[1][]{\ifthenelse{\equal{#1}{}}     
                                    {\rnkl{Z(\holgr_\qa)}{\LG}}        
                                   {\rnkl{Z(\holgr_{#1})}{\LG}}}       
\newcommand{\nkla}[1][]{\ifthenelse{\equal{#1}{}}      
                                    {\rnkl{\bz(\qa)}{\Gb}}        
                                    {\rnkl{\bz(#1)}{\Gb}}}       

\newcommand{\charakt}{\chi}                            

\newcommand{\YM}{{\text{YM}}}                          

\newcommand{\ymwirk}[1][]{\ifthenelse{\equal{#1}{}}{S_{\YM}}{S_{\YM,#1}}}

\newcommand{\bmat}{\begin{pmatrix}}
\newcommand{\emat}{\end{pmatrix}}

\newcommand{\ListNullAbstaende}{\setlength{\topsep}{0pt}%
                                \setlength{\parskip}{0pt}%
                                \setlength{\partopsep}{0pt}%
                                \setlength{\itemsep}{0pt}%
                                \setlength{\parsep}{0pt}}
\newcommand{\ListNurAnstrichAbstand}{\setlength{\topsep}{0pt}%
                                     \setlength{\parskip}{0pt}%
                                     \setlength{\partopsep}{0pt}%
                                     \setlength{\parsep}{0pt}}
\newenvironment{StandardListe}[2]%
               {\begin{list}%
                      {#1}%
                      {\settowidth{\leftmargin}{M#1}%
                       \settowidth{\labelwidth}{#1}%
                       \settowidth{\labelsep}{M}%
                       #2%
                      }%
                }%
               {\end{list}}%
\newenvironment{EinfachListe}[1]%
               {\begin{StandardListe}{#1}{\ListNullAbstaende}}%
               {\end{StandardListe}}%
               {\begin{StandardListe}{#1}{\ListNurAnstrichAbstand}}%
               {\end{StandardListe}}%
\newcommand{\labelsatz}[1]{#1}
\newcounter{listennr}                      %
\newlength{\hilfslaenge}
\newlength{\stdlabellaenge}
\newlength{\maximum}
\newcommand{\stdlabel}{}
\newcommand{\Maximum}{}
\newcommand{\iitem}[1][]{\ifthenelse{\equal{#1}{}}%
                           {\item \setlength{\hilfslaenge}{\stdlabellaenge}}%
                           {\item[\labelsatz{#1}\hfill]%
                            \settowidth{\hilfslaenge}{\labelsatz{#1}}}%
                         \ifthenelse{\lengthtest{\maximum < \hilfslaenge}}%
                           {\setlength{\maximum}{\hilfslaenge}%
                            \ifthenelse{\equal{#1}{}}%
                               {\renewcommand{\Maximum}{\stdlabel}}%
                               {\renewcommand{\Maximum}{#1}}}%
                           {}%
                      }      
\makeatletter
\newenvironment{AutoLabelLaengenListe}[2][]%
               {\begin{list}%
                      {\labelsatz{#1}\hfill}%
                      {\stepcounter{listennr}%
                       \settowidth{\leftmargin}{M\labelsatz{\ref{listnr\arabic{listennr}}}}%
                       \settowidth{\labelwidth}{\labelsatz{\ref{listnr\arabic{listennr}}}}%
                       \settowidth{\labelsep}{M}%
                       \settowidth{\stdlabellaenge}{\labelsatz{#1}}%
                       \renewcommand{\stdlabel}{#1}%
                       #2%
                       \renewcommand{\Maximum}{}%
                      }%
                }%
               {\renewcommand{\@currentlabel}{\Maximum}%
                \label{listnr\arabic{listennr}}%
                \end{list}%
                }%
\makeatother
\newenvironment{StandardEinrueckung}[2]%
               {\begin{list}%
                      {#1}%
                      {\settowidth{\leftmargin}{M#1}%
                       \settowidth{\labelwidth}{#1}%
                       \settowidth{\labelsep}{M}%
                       #2%
                      }%
                \item}%
               {\end{list}}%
\newenvironment{Einrueckungpur}[1]%
               {\begin{StandardEinrueckung}{#1}{\ListNullAbstaende}}%
               {\end{StandardEinrueckung}}%
\newenvironment{Einrueckung}[1]%
               {\begin{StandardEinrueckung}{#1}{\setlength{\parsep}{0pt}}}%
               {\end{StandardEinrueckung}}%

\makeatletter
\newcommand{\EineNumZeileGleichung}[2][0.5ex]
           {
            
            \vspace{#1} 
            \noindent
            \stepcounter{equation}
            \renewcommand{\@currentlabel}{\arabic{equation}}%
            \phantom{(\arabic{equation})}\hspace*{\fill}
            $\displaystyle{#2}$
            \hspace*{\fill}
            (\arabic{equation})

            \vspace{#1} 
            
           }
\makeatother
\makeatletter
\newcommand{\EineErwNumZeileGleichung}[2][0.5ex]
           {
            
            \vspace{#1} 
            \noindent
            \stepcounter{equation}
            \renewcommand{\@currentlabel}{\arabic{equation}}%
            \phantom{(\arabic{equation})}\hspace*{\fill}
            #2 %
            \hspace*{\fill}
            (\arabic{equation})

            \vspace{#1} 
            
           }
\makeatother
\newcommand{\breitrel}[1]{\hspace*{\tabcolsep} #1 \hspace*{\tabcolsep}}
\newlength{\abstaug}              %
\newenvironment{AllgUnnumGleichung}[2][1.0ex]
               {
  
                \setlength{\abstaug}{#1}
                \vspace{\abstaug}
                \hspace*{\fill}
                $\begin{array}[t]{#2}
                }%
               {\end{array}$
                \hspace*{\fill}
  
                \vspace{\abstaug}

                }%
\newenvironment{AllgNumGleichung}[2][0.0ex]
               {
  
                \setlength{\abstaug}{#1}
                \vspace{\abstaug}
                $\begin{tabular*}{\textwidth}[t]{#2}
                }%
               {\end{tabular*}$

                \vspace{\abstaug}

               }%
\newenvironment{StandardUnnumGleichungKlein}[1][0ex]
               {\renewcommand{\s}{\\[#1] }%
                \begin{AllgUnnumGleichung}{rcl}}%
               {\end{AllgUnnumGleichung}}%
\newenvironment{StandardUnnumGleichung}[1][0ex]%
               {\renewcommand{\s}{\\[#1] }%
                \begin{AllgUnnumGleichung}{>{\displaystyle}rc>{\displaystyle}l}}%
               {\end{AllgUnnumGleichung}}%
\newenvironment{XrelYZNumGleichung}[1][0ex]
               {\renewcommand{\s}{\\[#1] }%
                \begin{AllgNumGleichung}{rcll}}%
               {\end{AllgNumGleichung}}%

\newcommand{\erllang}[2][0.5\textwidth]%
              {\hfill\hspace*{1.5em}%
               \begin{minipage}[t]{#1}{\small%
                          \begin{list}{(}{\ListNullAbstaende%
                                          \settowidth{\leftmargin}{(}%
                                          \settowidth{\labelwidth}{(}%
                                          \settowidth{\labelsep}{}%
                                         }%
                          \item#2)%
                          \end{list}}%
               \end{minipage}\\[-0.9ex]
              }%
\newcommand{\DefBemUmgeb}[1]%
           {\newenvironment{#1}[1][]%
                           {\begin{Einrueckung}{{\bf #1}}%
                            \ifx##1\empty\else{{\bf ##1}
                            
                                                        }\fi%
                            }%
                           {\end{Einrueckung}}}
\newcommand{\DefSBemUmgeb}[2]
           {\newenvironment{#1}[1][]%
                           {\begin{Einrueckung}{{\bf #2}}%
                            \ifx##1\empty\else{{\bf ##1}
                            
                                                        }\fi%
                            }%
                           {\end{Einrueckung}}}
\makeatletter
\newcommand{\DefBspUmgeb}[3]
           {\newcounter{#2}[#3]%
            \newenvironment{#1}[1][]%
                           {\stepcounter{#2}%
                            \renewcommand{\ZaehlerMarke}{\arabic{#2}}%
                            \renewcommand{\Einzugsname}{{\bf #1 \ZaehlerMarke}}%
                            \begin{Einrueckung}{\Einzugsname}
                            \ifx##1\empty\else{{\bf ##1}\\}\fi%
                            \renewcommand{\@currentlabel}{\ZaehlerMarke}%
                            }%
                           {\end{Einrueckung}}}
\makeatother
\newcommand{\ZaehlerbisEbene}{section}
\newcommand{\Ebenea}{section}
\newcommand{\Ebeneb}{subsection}

\newcommand{\Abschnittnummer}{%
            \ifx\ZaehlerbisEbene\Ebenea{\arabic{section}}%
             \else{%
              \ifx\ZaehlerbisEbene\Ebeneb{\arabic{section}.\arabic{subsection}}%
               \else{\arabic{section}.\arabic{subsection}.\arabic{subsubsection}}%
              \fi}%
            \fi}     
\newcommand{\Abschnittnummerpunkt}{\Abschnittnummer.}     
\newcommand{\Einzugsname}{}
\newcommand{\ZaehlerMarke}{}
\makeatletter
\newcommand{\DefThmUmgeb}[3]%
           {\newcounter{#1}[#3]%
            \newenvironment{#1}[1][]%
                           {\stepcounter{#2}%
                            \setcounter{#1}{\value{#2}}%
                            \renewcommand{\ZaehlerMarke}{\Abschnittnummerpunkt\arabic{#1}}%
                            \renewcommand{\Einzugsname}{{\bf #1 \ZaehlerMarke}}%
                            \begin{Einrueckung}{\Einzugsname}
                            \ifx##1\empty\else{{\bf ##1}
                            
                                                        }\fi%
                            \renewcommand{\@currentlabel}{\ZaehlerMarke}%
                            }%
                           {\end{Einrueckung}}}
\makeatother
\makeatletter
\newcommand{\DefSThmUmgeb}[4]%
           {\newcounter{#1}[#3]%
            \newenvironment{#1}[1][]%
                           {\stepcounter{#2}%
                            \setcounter{#1}{\value{#2}}%
                            \renewcommand{\ZaehlerMarke}{\Abschnittnummerpunkt\arabic{#1}}%
                            \renewcommand{\Einzugsname}{{\bf #4 \ZaehlerMarke}}
                            \begin{Einrueckung}{\Einzugsname}
                            \ifx##1\empty\else{{\bf ##1}

                                                        }\fi%
                            \renewcommand{\@currentlabel}{\ZaehlerMarke}%
                            }%
                           {\end{Einrueckung}}}
\makeatother
\makeatletter
\newcommand{\DefUnterNumThmUmgeb}[5]%
           {\newcounter{#1}[#3]%
            \newcounter{#4}%
            \newenvironment{#1}[1][]%
                           {\ifx##1\empty\else{\stepcounter{#2}\setcounter{#4}{0}}\fi%
                            \stepcounter{#4}%
                            \setcounter{#1}{\value{#2}}%
                            \renewcommand{\ZaehlerMarke}{\Abschnittnummerpunkt\arabic{#1}\alph{#4}}%
                            \renewcommand{\Einzugsname}{{\bf #5 \ZaehlerMarke}}
                            \begin{Einrueckung}{\Einzugsname}
                            \renewcommand{\@currentlabel}{\ZaehlerMarke}%
                            }%
                           {\end{Einrueckung}}}
\makeatother
\newenvironment{Beweis}[1][]%
               {\begin{Einrueckung}{{\bf Beweis}}%
                \ifx#1\empty\else{{\bf #1}

                                            }\fi%
                }%
               {\end{Einrueckung}%
                }%
\newenvironment{Proof}[1][]%
               {\begin{Einrueckung}{{\bf Proof}}%
                \ifx#1\empty\else{{\bf #1}

                                            }\fi%
                }%
               {\end{Einrueckung}%
                }%
               {\begin{Einrueckung}{{\bf \glqq Beweis\grqq}}%
                \ifx#1\empty\else{{\bf #1}
                
                                            }\fi%
                }%
               {\end{Einrueckung}%
                }%
               {\begin{Einrueckung}{{\bf Begr"undung}}%
                \ifx#1\empty\else{{\bf #1}
                
                                            }\fi%
                }%
               {\end{Einrueckung}%
                }%
\newenvironment{Hinrichtung}%
               {\begin{Einrueckungpur}{$\impliz$}}%
               {\end{Einrueckungpur}}%
\newenvironment{Rueckrichtung}%
               {\begin{Einrueckungpur}{$\invimpliz$}}%
               {\end{Einrueckungpur}}%
               {\begin{Einrueckungpur}{\glqq$\teilmenge$\grqq}}%
               {\end{Einrueckungpur}}%
               {\begin{Einrueckungpur}{\glqq$\obermenge$\grqq}}%
               {\end{Einrueckungpur}}%
               {\begin{Einrueckungpur}{"$\teilmenge$"}}%
               {\end{Einrueckungpur}}%
               {\begin{Einrueckungpur}{"$\obermenge$"}}%
               {\end{Einrueckungpur}}%
\newcommand{\qed}{\nopagebreak\hspace*{2em}\hspace*{\fill}{\bf qed}}
\newcommand{\ARabic}{\arabic}
\newcommand{\Nummerntypa}{\arabic}   
\newcommand{\Nummerntypb}{\alph}
\newcommand{\Nummerntypc}{\roman}
\newcommand{\Nummerntypd}{\Alph}

\newcommand{\Nra}{\Nummerntypa{Nummera}}            %
\newcommand{\Nrb}{\Nummerntypb{Nummerb}}            %
\newcommand{\Nrc}{\Nummerntypc{Nummerc}}                
\newcommand{\Nrd}{\Nummerntypd{Nummerd}}                
\newcommand{\ZeichenzuNrTyp}[1]%
           {\ifx#1\ARabic {.}\else{)}%
                  \fi}                              %
\newcommand{\NrZeicha}{\ZeichenzuNrTyp{\Nummerntypa}}
\newcommand{\NrZeichb}{\ZeichenzuNrTyp{\Nummerntypb}}
\newcommand{\NrZeichc}{\ZeichenzuNrTyp{\Nummerntypc}}
\newcommand{\NrZeichd}{\ZeichenzuNrTyp{\Nummerntypd}}
\newcommand{\ListMarkea}%
           {\Nra\NrZeicha}
\newcommand{\ListMarkeb}%
           {\Nra\NrZeicha\Nrb\NrZeichb}
\newcommand{\ListMarkec}%
           {\Nra\NrZeicha\Nrb\NrZeichb\Nrc\NrZeichc}
\newcommand{\ListMarked}%
           {\Nra\NrZeicha\Nrb\NrZeichb\Nrc\NrZeichc\Nrd\NrZeichd}
\newcommand{\Anfangszeichen}{}
\newcommand{\Anfangspunkt}{}
\newcounter{Schachtelebene}
\newcounter{Hilfszaehler}
\newcommand{\Hilfsbefehl}{}
\newcommand{\Schachtelebene}{\alph{Schachtelebene}}
\makeatletter
\newenvironment{AllgNumerierteListe}[2][]
               {\addtocounter{Schachtelebene}{1}%
		\setcounter{Hilfszaehler}{#2}%
                \renewcommand{\Anfangszeichen}%
                             {\renewcommand{\Hilfsbefehl}{\csname Nummerntyp\Schachtelebene \endcsname}%
                              \Hilfsbefehl{Hilfszaehler}}%
                \renewcommand{\Anfangspunkt}%
                             {\csname NrZeich\Schachtelebene \endcsname}%
                \begin{list}%
                      {\stepcounter{Nummer\Schachtelebene}%
                       \csname Nr\Schachtelebene \endcsname
                       \csname NrZeich\Schachtelebene \endcsname
                       }%
                      {\settowidth{\leftmargin}{M\Anfangszeichen\Anfangspunkt}%
                       \settowidth{\labelwidth}{\Anfangszeichen\Anfangspunkt}%
                       \settowidth{\labelsep}{M}%
                       \setlength{\topsep}{0pt}%
                       \setlength{\parskip}{0pt}%
                       \setlength{\partopsep}{0pt}%
                       \setlength{\itemsep}{0pt}%
                       \setlength{\parsep}{0pt}%
                      }%
                \renewcommand{\@currentlabel}{\csname ListMarke\Schachtelebene \endcsname}%
                }%
               {\ifthenelse{\equal{}{}}{\setcounter{Nummer\Schachtelebene}{0}}{}
                \addtocounter{Schachtelebene}{-1}%
                \end{list}}
\makeatother
\newenvironment{NumerierteListe}[1]
               {\begin{AllgNumerierteListe}{#1}}
               {\end{AllgNumerierteListe}}
\newenvironment{WeiterNumerierteListe}[1]
               {\begin{AllgNumerierteListe}[Weiter]{#1}}
               {\end{AllgNumerierteListe}}

\newcommand{\UnnumAnfangszeichen}{}
\newcounter{UnnumSchachtelebene}
\newcommand{\UnnumSchachtelebene}{\alph{UnnumSchachtelebene}}
\makeatletter
\newenvironment{UnnumerierteListe}%
               {\addtocounter{UnnumSchachtelebene}{1}%
                \renewcommand{\UnnumAnfangszeichen}%
                             {\csname UnnumZeich\UnnumSchachtelebene \endcsname}%
                \begin{list}%
                      {\UnnumAnfangszeichen}%
                      {\settowidth{\leftmargin}{M\UnnumAnfangszeichen}%
                       \settowidth{\labelwidth}{\UnnumAnfangszeichen}%
                       \settowidth{\labelsep}{M}%
                       \setlength{\topsep}{0pt}%
                       \setlength{\parskip}{0pt}%
                       \setlength{\partopsep}{0pt}%
                       \setlength{\itemsep}{0pt}%
                       \setlength{\parsep}{0pt}%
                      }%
                }%
               {\addtocounter{UnnumSchachtelebene}{-1}%
                \end{list}}
\makeatother
\newlength{\fktdefhilfslaenge}
\newcommand{\ohnefktdef}[4]
           {\hspace*{\fill}
            $\begin{array}[t]{ccc}%
            #1 & \nach & #2 \\
            #3 & \auf  & #4
            \end{array}$
            \hspace*{\fill}}
\newcommand{\fktdef}[5]
           {\hspace*{\fill}
            $\begin{array}[t]{cccc}%
            #1: & #2 & \nach & #3 \\    
                & #4 & \auf  & #5
            \end{array}$
            \settowidth{\fktdefhilfslaenge}{$#1$:}
            \hspace*{0.6 \fktdefhilfslaenge}  
            \hspace*{\fill}}
\newcommand{\fktdefpur}[5]
           {$\begin{array}[t]{cccc}%
            #1: & #2 & \nach & #3 \\    
                & #4 & \auf  & #5
            \end{array}$}
\newcommand{\fktdefabgesetztpur}[5]
           {
            
            $\begin{array}[t]{cccc}%
            #1: & #2 & \nach & #3 \\    
                & #4 & \auf  & #5
            \end{array}$
            \settowidth{\fktdefhilfslaenge}{$#1$:}
            \hspace*{0.6 \fktdefhilfslaenge}
            
           }
\newcommand{\fktdefabgesetzt}[5]
           {
           
            \hspace*{\fill}
            $\begin{array}[t]{cccc}%
            #1: & #2 & \nach & #3 \\    
                & #4 & \auf  & #5
            \end{array}$
            \settowidth{\fktdefhilfslaenge}{$#1$:}
            \hspace*{0.6 \fktdefhilfslaenge}  
            \hspace*{\fill}
            
            }
\newcommand{\ohnefktdefabgesetzt}[4]
           {      

            \hspace*{\fill}
            $\begin{array}[t]{ccc}%
            #1 & \nach & #2 \\
            #3 & \auf  & #4
            \end{array}$
            \hspace*{\fill}

            }
\newcommand{\doppelohnefktdefabgesetzt}[6]
           {

            \hspace*{\fill}
            $\begin{array}[t]{ccccc}%
            #1 & \nach & #2 & \nach & #3\\
            #4 & \auf  & #5 & \auf  & #6
            \end{array}$
            \hspace*{\fill}

            }
\newcommand{\anhang}%
           {\appendix
            \sectioninh{Anhang}
            \renewcommand{\Abschnittnummer}{%
                  \ifx\ZaehlerbisEbene\Ebenea{\Alph{section}}%
                  \else{%
                        \ifx\ZaehlerbisEbene\Ebeneb{\Alph{section}.\arabic{subsection}}%
                        \else{\Alph{section}.\arabic{subsection}.\arabic{subsubsection}}%
                        \fi}%
                  \fi}%
            \renewcommand{\Abschnittnummerpunkt}{\Abschnittnummer.}     
            }            
\newcommand{\anhangengl}%
           {\appendix
            \sectioninh{Appendix}
            \renewcommand{\Abschnittnummer}{%
                  \ifx\ZaehlerbisEbene\Ebenea{\Alph{section}}%
                  \else{%
                        \ifx\ZaehlerbisEbene\Ebeneb{\Alph{section}.\arabic{subsection}}%
                        \else{\Alph{section}.\arabic{subsection}.\arabic{subsubsection}}%
                        \fi}%
                  \fi}%
            \renewcommand{\Abschnittnummerpunkt}{\Abschnittnummer.}     
            }

\newcounter{wdhlstufe}
\newcommand{\sectioninh}[1]%
           {\section*{#1}%
            \addcontentsline{toc}{section}{#1}}
\newcommand{\bezeichnung}[3]%
           {\begin{Einrueckungpur}{\hbox to 6em{#1}\hbox to 2.4em{\hfill#2}}
            #3
            \end{Einrueckungpur}}

\newcommand{\doppelteinfach}{e}

\newcommand{\ifdoppelt}[1]{\ifthenelse{\equal{\doppelteinfach}{d}}{#1}{}}
\newcommand{\ifeinfach}[1]{\ifthenelse{\equal{\doppelteinfach}{e}}{#1}{}}

\newlength{\querfhilfsl}              %

\newlength{\hll}

\DefThmUmgeb{Theorem}{Theorem}{\ZaehlerbisEbene}
\DefThmUmgeb{Definition}{Definition}{\ZaehlerbisEbene}
\DefThmUmgeb{Satz}{Theorem}{\ZaehlerbisEbene}
\DefThmUmgeb{Proposition}{Theorem}{\ZaehlerbisEbene}
\DefThmUmgeb{Lemma}{Theorem}{\ZaehlerbisEbene}
\DefThmUmgeb{Folgerung}{Theorem}{\ZaehlerbisEbene}
\DefThmUmgeb{Corollary}{Theorem}{\ZaehlerbisEbene}
\DefThmUmgeb{Vorschrift}{Definition}{\ZaehlerbisEbene}
\DefThmUmgeb{Construction}{Definition}{\ZaehlerbisEbene}
\DefSThmUmgeb{FormSatz}{Theorem}{\ZaehlerbisEbene}{\glqq Satz\grqq} 
\DefThmUmgeb{Vermutung}{Theorem}{\ZaehlerbisEbene}
\DefThmUmgeb{Conjecture}{Theorem}{\ZaehlerbisEbene}
\DefThmUmgeb{Konvention}{Definition}{\ZaehlerbisEbene}
\DefThmUmgeb{Feststellung}{Theorem}{\ZaehlerbisEbene}
\DefUnterNumThmUmgeb{DefinitionZusatzNum}{Definition}{\ZaehlerbisEbene}{DefZN}{Definition}
\DefBspUmgeb{Beispiel}{Beispiel}{subsubsection}
\DefBspUmgeb{Example}{Example}{subsubsection}
\DefBspUmgeb{Frage}{Frage}{section}
\DefBspUmgeb{Question}{Question}{section}
\DefBspUmgeb{Aufgabe}{Aufgabe}{section}
\DefBspUmgeb{Ziel}{Ziel}{section}
\DefBemUmgeb{Bemerkung}
\DefBemUmgeb{Remark}
\DefSBemUmgeb{OffeneFrage}{Offene Frage}
\newcommand{\bdf}{\begin{Definition}}
\newcommand{\edf}{\end{Definition}}
\newcommand{\bvorsch}{\begin{Vorschrift}}
\newcommand{\evorsch}{\end{Vorschrift}}
\newcommand{\bconst}{\begin{Construction}}
\newcommand{\econst}{\end{Construction}}
\newcommand{\bthm}{\begin{Theorem}}
\newcommand{\ethm}{\end{Theorem}}
\newcommand{\bsatz}{\begin{Satz}}
\newcommand{\esatz}{\end{Satz}}
\newcommand{\bprop}{\begin{Proposition}}
\newcommand{\eprop}{\end{Proposition}}
\newcommand{\blem}{\begin{Lemma}}
\newcommand{\elem}{\end{Lemma}}
\newcommand{\bfolg}{\begin{Folgerung}}
\newcommand{\efolg}{\end{Folgerung}}
\newcommand{\bcorr}{\begin{Corollary}}
\newcommand{\ecorr}{\end{Corollary}}
\newcommand{\bfest}{\begin{Feststellung}}
\newcommand{\efest}{\end{Feststellung}}
\newcommand{\bbew}{\begin{Beweis}}
\newcommand{\ebew}{\end{Beweis}}
\newcommand{\bpf}{\begin{Proof}}
\newcommand{\epf}{\end{Proof}}
\newcommand{\bwnum}{\begin{WeiterNumerierteListe}}
\newcommand{\ewnum}{\end{WeiterNumerierteListe}}
\newcommand{\bdfzn}{\begin{DefinitionZusatzNum}}
\newcommand{\edfzn}{\end{DefinitionZusatzNum}}
\newcommand{\bbem}{\begin{Bemerkung}}
\newcommand{\ebem}{\end{Bemerkung}}
\newcommand{\brem}{\begin{Remark}}
\newcommand{\erem}{\end{Remark}}
\newcommand{\bnum}{\begin{NumerierteListe}}
\newcommand{\enum}{\end{NumerierteListe}}
\newcommand{\bunum}{\begin{UnnumerierteListe}}
\newcommand{\eunum}{\end{UnnumerierteListe}}
\newcommand{\bbsp}{\begin{Beispiel}}
\newcommand{\ebsp}{\end{Beispiel}}
\newcommand{\bex}{\begin{Example}}
\newcommand{\eex}{\end{Example}}
\newcommand{\bfrag}{\begin{Frage}}
\newcommand{\efrag}{\end{Frage}}
\newcommand{\bquest}{\begin{Question}}
\newcommand{\equest}{\end{Question}}
\newcommand{\baufg}{\begin{Aufgabe}}
\newcommand{\eaufg}{\end{Aufgabe}}
\newcommand{\bof}{\begin{OffeneFrage}}
\newcommand{\eof}{\end{OffeneFrage}}
\newcommand{\bverm}{\begin{Vermutung}}
\newcommand{\everm}{\end{Vermutung}}
\newcommand{\bconj}{\begin{Conjecture}}
\newcommand{\econj}{\end{Conjecture}}
\newcommand{\bkonv}{\begin{Konvention}}
\newcommand{\ekonv}{\end{Konvention}}
\newcommand{\bglklein}{\begin{StandardUnnumGleichungKlein}}
\newcommand{\eglklein}{\end{StandardUnnumGleichungKlein}}
\newcommand{\bgl}{\begin{StandardUnnumGleichung}}
\newcommand{\egl}{\end{StandardUnnumGleichung}}
\newcommand{\bglrtext}{\begin{XrelYZNumGleichung}}
\newcommand{\eglrtext}{\end{XrelYZNumGleichung}}

\newcommand{\berlgl}{\begin{StandardUnnumGleichung}}
\newcommand{\eerlgl}{\end{StandardUnnumGleichung}}
\newcommand{\beinrueck}{\begin{Einrueckungpur}} 
\newcommand{\eeinrueck}{\end{Einrueckungpur}}
\newcommand{\beinflist}{\begin{EinfachListe}} 
\newcommand{\eeinflist}{\end{EinfachListe}}
\newcommand{\beq}{\begin{equation}}
\newcommand{\eeq}{\end{equation}}
\newcommand{\bhin}{\begin{Hinrichtung}}
\newcommand{\ehin}{\end{Hinrichtung}}
\newcommand{\brueck}{\begin{Rueckrichtung}}
\newcommand{\erueck}{\end{Rueckrichtung}}
\newcommand{\bvl}{\begin{AutoLabelLaengenListe}{\ListNullAbstaende}}
\newcommand{\evl}{\end{AutoLabelLaengenListe}}
\newcommand{\df}[1]{{\bf #1}}

\newlength{\adressabstand}
\DefThmUmgeb{Convention}{Definition}{\ZaehlerbisEbene}
\DefThmUmgeb{Notation}{Definition}{\ZaehlerbisEbene}
\makeatletter
\newenvironment{AllgNumerierteListeWeiter}[2][]
               {\addtocounter{Schachtelebene}{1}%
		\setcounter{Hilfszaehler}{#2}%
                \renewcommand{\Anfangszeichen}%
                             {\renewcommand{\Hilfsbefehl}{\csname Nummerntyp\Schachtelebene \endcsname}%
                              \Hilfsbefehl{Hilfszaehler}}%
                \renewcommand{\Anfangspunkt}%
                             {\csname NrZeich\Schachtelebene \endcsname}%
                \begin{list}%
                      {\stepcounter{Nummer\Schachtelebene}%
                       \csname Nr\Schachtelebene \endcsname
                       \csname NrZeich\Schachtelebene \endcsname
                       }%
                      {\settowidth{\leftmargin}{M\Anfangszeichen\Anfangspunkt}%
                       \settowidth{\labelwidth}{\Anfangszeichen\Anfangspunkt}%
                       \settowidth{\labelsep}{M}%
                       \setlength{\topsep}{0pt}%
                       \setlength{\parskip}{0pt}%
                       \setlength{\partopsep}{0pt}%
                       \setlength{\itemsep}{0pt}%
                       \setlength{\parsep}{0pt}%
                      }%
                \renewcommand{\@currentlabel}{\csname ListMarke\Schachtelebene \endcsname}%
                }%
               {\addtocounter{Schachtelebene}{-1}%
                \end{list}}
\makeatother
\renewcommand{\bwnum}{\begin{AllgNumerierteListeWeiter}}
\renewcommand{\ewnum}{\end{AllgNumerierteListeWeiter}}
\renewcommand{\boundfkt}{C_b}

\newcommand{\Bigbetrag}[1]%
           {\Bigl|{#1}\Bigr|}            
\newcommand{\Biggbetrag}[1]%
           {\Biggl|{#1}\Biggr|}            

\newcommand{\Bignorm}[2][]{\Bigl\lVert#2\Bigr\rVert_{#1}}

\newcommand{\Bigsupnorm}[1]{\Bignorm[\infty]{#1}}

\newcommand{\bconv}{\begin{Convention}}
\newcommand{\econv}{\end{Convention}}
\newcommand{\bnot}{\begin{Notation}}
\newcommand{\enot}{\end{Notation}}

\DeclareMathOperator{\realteil}{Re}
\DeclareMathOperator{\imaginaerteil}{Im}

\renewcommand{\Im}{\imaginaerteil}
\renewcommand{\re}{\realteil}

\newcommand{\bpm}{\begin{pmatrix}}
\newcommand{\epm}{\end{pmatrix}}

\DeclareMathOperator{\spec}{spec}

\newcommand{\linf}{\ell^\infty}

\begin{document}
\title{Continuity of States on Non-Unital Differential Algebras in Loop Quantum Cosmology}
\author{Christian Fleischhack\thanks{e-mail: 
            {\tt fleischh@math.upb.de}} \\   
        \\
        {\normalsize\em Institut f\"ur Mathematik}\\[\adressabstand]
        {\normalsize\em Universit\"at Paderborn}\\[\adressabstand]
        {\normalsize\em Warburger Stra\ss e 100}\\[\adressabstand]
        {\normalsize\em 33098 Paderborn}\\[\adressabstand]
        {\normalsize\em Germany}
        \\[-25\adressabstand]}     
\date{March 23, 2018}
\maketitle
\newcommand{\iotarestr}{\iota_\stdrestr}
\newcommand{\stdrestr}{\setabb}
\newcommand{\algrestr}[1][\stdrestr]{\alg_{#1}}
\newcommand{\blgrestr}[1][\stdrestr]{\blg_{#1}}
\newcommand{\clg}{{\mathfrak C}}
\newcommand{\dlg}{{\mathfrak D}}
\newcommand{\elg}{{\mathfrak E}}
\newcommand{\tlg}{{\mathfrak T}}
\newcommand{\urbildalg}{\dlg}
\newcommand{\fktseteins}{{\mathfrak E}}
\newcommand{\fktsetzwei}{{\mathfrak F}}
\newcommand{\ptfktset}{\dlg}
\newcommand{\set}{\mathbf S}
\newcommand{\topset}{\set}   
\newcommand{\elset}{s} 
\renewcommand{\malg}{\spec \alg} 
\newcommand{\maltalg}{\spec \altalg} 
\newcommand{\Y}{\mathbf{Y}} 
\newcommand{\ely}{\mathbf{y}} 
\newcommand{\restr}[1]{#1_{\stdrestr}}
\newcommand{\elalg}{a} 
\newcommand{\elblg}{b} 
\newcommand{\elclg}{c} 
\newcommand{\eldlg}{d} 
\newcommand{\elelg}{e} 
\newcommand{\elurbildalg}{\eldlg} 
\newcommand{\elfktseteins}{e}
\newcommand{\elfktsetzwei}{f}
\newcommand{\mclg}{\spec \clg} 
\newcommand{\charabb}{\tau}
\newcommand{\charact}{\charakt}
\newcommand{\altcharakt}{\phi}
\newcommand{\disjunion}{\sqcup}
\newcommand{\setabb}{\sigma}
\newcommand{\redeinbett}{\lambda}
\newcommand{\altredeinbett}{\mu}
\newcommand{\plus}{p}
\newcommand{\cover}{{\cal U}}
\renewcommand{\linf}{\ell^\infty}
\newcommand{\gelf}[1][]{\ifthenelse{\equal{#1}{}}{\widetilde}{G_{#1}}}
\newcommand{\gelftrf}{{\sim}}
\newcommand{\coverbasis}{\cover_{\text{Basis}}}
\newcommand{\lin}{{\text{lin}}}
\newcommand{\gelflang}[1]{(#1)^\gelftrf}
\renewcommand{\gelftrf}{{\sim}}
\newcommand{\setabbalt}{\tau}
\newcommand{\action}{\varphi}
\newcommand{\invwegaction}[2]{\action_{#1}^{-1}(#2)}
\renewcommand{\invwegaction}[2]{#1^{-1}#2}
\newcommand{\wegaction}[2]{\action_{#1}(#2)}
\renewcommand{\wegaction}[2]{#1#2}
\renewcommand{\set}{X}
\newcommand{\config}{\mathcal{C}}
\renewcommand{\set}{\mathbf S}
\renewcommand{\topset}{\set}   
\renewcommand{\elset}{\mathbf s} 
\renewcommand{\set}{\mathcal X}
\newcommand{\elaltaltset}{z}
\newcommand{\altaltset}{\mathcal Z}
\newcommand{\altset}{\mathcal Y}
\newcommand{\altalg}{\blg}
\newcommand{\caneinbett}{\varkappa}
\renewcommand{\elset}{x}
\newcommand{\elaltset}{y}
\renewcommand{\config}{\set}
\newcommand{\alt}[1]{\emph{{\small#1}}}
\newcommand{\natmap}{\iota}
\newcommand{\altnatmap}{\iota_\altset}
\newcommand{\kleindlg}{\mathfrak d}
\newcommand{\sfkt}{f}
\newcommand{\sfktalt}{g}
\newcommand{\sfktaltalt}{h}
\newcommand{\dsa}{\mathcal D}

\begin{abstract}
In a recent paper \cite{d117}, Engle, Hanusch and Thiemann 
showed that there is a unique state on the
reduced holonomy-flux $\ast$-algebra of 
homogeneous isotropic loop quantum
cosmology,
that is 
invariant under residual diffeomorphims. 
This result has been claimed to be true 
both for the Ashtekar-Bojowald-Lewandowski framework
and for that introduced by the present author. 
Unfortunately, the uniqueness proof relies on an incorrect argument which 
spoils the second case. In our short note, we are going to
patch this issue, this way keeping the nice uniqueness result in both 
cases. Moreover, we will even extend the underlying operator algebraic statements
as this might help later for
studying higher-dimensional models.
\end{abstract}

\enlargethispage{-0.2\baselineskip}

\section{Introduction}
Representation theory 
has turned out indispensable
for many mathematically rigorous quantum theories. 
Particularly strong statements come from uniqueness results
like the celebrated Stone-von Neumann theorem, 
giving uniqueness in quantum mechanics, 
or the recent results in loop quantum gravity 
on the holonomy-flux \cite{lost} as well as on the Weyl algebra \cite{paper39}.
So it comes with no surprise that one is looking for their counterparts
also in the realm of loop quantum cosmology. 
Indeed, Engle, Hanusch and Thiemann have recently claimed \cite{d117} that there is
a unique invariant state on the holonomy-flux $\ast$-algebra 
also for homogeneous isotropic cosmologies. Here, invariance is understood
w.r.t.\ so-called residual diffeomorphisms, i.e., those diffeomorphisms
that do not destroy the symmetry (here: homogeneity and isotropy, taking their action
on the fiducial cell into account). Although we believe that their uniqueness 
result is correct,
its proof contains a flaw concerning the continuity of some state.
To explain the problem, let us consider a state $\omega$ on some 
$\ast$-subalgebra $\dlg$ of some abelian $C^\ast$-algebra $\clg = \boundfkt(X)$. 
To prove continuity of $\omega$, Engle et al.\ used that it is sufficient to show 
(see below) that 
$\sqrt{1 + \varphi}$ is in $\dlg$  
for each real-valued $\varphi \in \dlg$ with $\supnorm\varphi < 1$. 
Unfortunately, $\sqrt{1 + \varphi} \in \dlg$ implies that $\dlg$ is unital.
Unitality, however, is not given for $\dlg = C_0(\R)$ which is the second case considered in \cite{d117}
and is needed to prove uniqueness for the embeddable loop quantum cosmology case.

Fortunately, it is not very difficult to modify this step in the 
proof without modifying the ultimate uniqueness claim as we will show
in this short notice. We will prove results that are somewhat more general 
than needed for just closing the gap in \cite{d117} 
as these extensions might become useful for investigations 
of models with more degrees of freedom like Bianchi I.

\section{Relation to the Engle-Hanusch-Thiemann Paper}

In the whole article, let there be 

\bgl
M & \ldots & \text{some open set in a Banach space} \\
\alg & \ldots & \text{some $\ast$-algebra} \\
\blg & \ldots & \text{some unital Banach $\ast$-algebra} \\
\clg & \ldots & \text{some Banach $\ast$-subalgebra of $\boundfkt(M,\blg)$} \\
\egl\noindent
Note that we assume the norm on $\blg$ to fulfill $\norm{\elblg^\ast} = \norm \elblg$ 
for all $\elblg \in \blg$. Moreover, $\boundfkt(M,\blg)$ denotes 
the set of bounded continuous functions from $M$ to $\blg$ and is equipped with the
usual supremum norm.
Finally, observe that 
our results comprise, in particular, the situation
\bgl
M & \ldots & \R \\
\alg & \ldots & \text{quantum reduced holonomy-flux $\ast$-algebra} \\
\blg & \ldots & \C \\
\clg & \ldots & \text{either $C_0(\R)$ or $C_\AP(\R)$ or $C_0(\R) \dirsum C_\AP(\R)$} \\
\egl\noindent
which is exactly the situation studied by 
Engle, Hanusch and Thiemann in \cite{d117}.

\section{States}

\bdf
A \df{state} on 
$\alg$ is a $\ast$-linear functional 
$\omega : \alg \nach \C$ which 
is \df{weakly positive}, 
i.e.\ it fulfills 
\bgl
\omega(\elalg^\ast \elalg) & \geq & 0 \qquad \text{for all $\elalg \in \alg$.}
\egl
\edf
For our purposes, we do not require a state to be normalized (i.e.\ to fulfill 
$\omega(\EINS) = 1$) as soon as $\alg$ is unital.

\section{Engle-Hanusch-Thiemann Strategy}
Let us recall the main argument in question (somewhat adapted to our notation). 
In Subsection ``Continuity''\ of Section 4 of \cite{d117}, Engle et al.\ 
considered $\kleindlg$ to contain those functions in 
$C_0(\R)$ or $C_\AP(\R)$
that are smooth and any of their derivatives are in 
$C_0(\R)$ or $C_\AP(\R)$, respectively, again.
Moreover, there is a $\ast$-homomorphism $I$ from $\kleindlg$ to the reduced holonomy-flux $\ast$-algebra
$\alg$ and a state $\omega$ on $\alg$. Now, the authors from \cite{d117} 
claim that this already implies that $\dach \omega := \omega \circ I$ is continuous.
The idea for the proof was to use that $\sqrt{1 \pm t}$ is analytic for $\betrag t < 1$, whence
$1 \pm \varphi$ for real-valued $\varphi$ on $\R$ can be written 
(either by functional calculus or even more direct by Taylor expansion)
as $\quer \psi  \psi$ with $\psi := \sqrt{\EINS \pm \varphi} \in \kleindlg$, provided $\supnorm \varphi < 1$.
Now, the state property implies
$\dach\omega(1 \pm \varphi) = \dach\omega(\quer \psi \psi) = \omega(I(\psi)^\ast I(\psi)) \geq 0$.
A straightforward argument shows now that $\dach\omega(\varphi)$ is at most $1$ for normalized
$\omega$.

At a first glance, the proof above is nice and complete. However, the existence of $\psi$ is
only given for $\quer\kleindlg$ being $C_\AP(\R)$, but not for $C_0(\R)$. In fact, in the latter case,
$\psi$ goes to $1$ at infinity, but not to $0$ as required. More abstractly, the
argument only goes through if $\kleindlg$ is unital itself. 
That unitality is indeed needed in general, will be shown in Section \ref{sect:ex-non-cont-state}
where we construct a non-continuous state in a non-unital example.

Nevertheless, fixing that issue in \cite{d117} is not very difficult. Indeed, in the situation 
of \cite{d117}, the \emph{overall}\/ framework is unital. More concretely, the
relevant commutative algebra (denoted by $\dlg$ there) 
is in both cases unital, namely 
$C_\AP(\R)$ or $C_0(\R) \dirsum C_\AP(\R)$, respectively, after completion.
The solution is now just to ``steal'' the unit from the upper level in order to get
the desired continuity. This will be done in the proof of Proposition \ref{prop:cont-crit-state}.

\section{Example of Non-Continuous State}
\label{sect:ex-non-cont-state}
Before going to do this, let us construct a non-continuous state on some 
normed $\ast$-algebra.

For this, let us consider the product space $X$ of countably many unit intervals $[0,1]$.
Then the canonical projections $x_n : X \nach [0,1]$ are, of course, 
bounded continuous
functions, i.e., $x_n \in \boundfkt(X)$.
Denote by $\dlg$ the $\ast$-subalgebra of $\boundfkt(X)$ generated by all these $x_n$.
It is clear that the monomials, i.e.\ the finite products of $x_n$ with repetitions admitted, 
form a vector space basis 
for $\dlg$.
Note that the unit function $\EINS$ is not contained in $\dlg$.
Now, we define $\omega : \dlg \nach \C$ to be the linear functional
that maps $x_n$ to $n$ and any monomial of degree $2$ or more to $0$.
Obviously, $\omega$ is linear. It is even a state. In fact,
if $f \in \dlg$, then $f^\ast f$ is a sum of monomials of degree at least $2$,
giving $\omega(f^\ast f) \ident 0$.
On the other hand, we see that $\supnorm{x_n} = 1$ for all $n$, but
$\omega(x_n) = n$, giving 
non-continuity of $\omega$.

\section{Differential Algebra}

Let us now come back to the general situation. We introduce

\bdf
The \df{differential algebra} $\dsa(\clg)$ of $\clg$ is given
by
\bgl
\dsa(\clg) 
 & := & \{\varphi \in \clg \mid \text{all partial derivatives (of any order) of $\varphi$ are in $\clg$}\}\!\!\!
\egl
\edf
Note that we consider partial derivatives w.r.t.\ subspaces of $M$ with real dimension $1$.
Moreover, $\varphi \in \dsa(\clg)$ tacitly includes the assumption that all the partial derivatives of $\varphi$ 
exist. Obviously, we have
\blem
$\dsa(\clg)$ is a $\ast$-subalgebra of $\clg$.
\elem

\brem
\bnum2
\item
For homogeneous isotropic cosmology, $\dsa(C_\AP(\R))$ and $\dsa(C_0(\R))$ 
correspond to $\kleindlg_\AP$ and $\kleindlg_0$, respectively, in \cite{d117}.
\item
For Bianchi I, the differential algebra 
$\dsa(C_\AP(\R^3))$ corresponds to $\dlg$ in Section 5 of \cite{d117} for the 
standard LQC configuration space.
As for the isotropic case, this differential algebra is dense in $\clg$.\ \cite{d117}

In the embeddable LQC version, however, the situation remains open.
At a first glance, this appears to be a surprise as also for Bianchi I, 
the holonomy algebra
$\clg$ is generated by smooth functions; indeed,
the underlying differential equation depends analytically 
on the parameters $c_1$, $c_2$ and $c_3$.
But, it has still been unknown whether the partial derivatives 
are in $\clg$ again.
Even worse: to the best of our knowledge, the explicit
form of $\clg$ has remained unknown (as already mentioned in \cite{d117}).

\enum

\erem

\section{Closedness of $\dsa(\clg)$ under Analytic Functions}

Extending the arguments of \cite{d117}, we get

\blem
\label{lem:closed-under-analytic}
Let $\sfkt$ be analytic at $0$ with convergence radius $r$ and $\sfkt(0) = 0$.

Then $\sfkt \circ \varphi \in \dsa(\clg)$
for all $\varphi \in \dsa(\clg)$ with $\supnorm\varphi < r$.
\elem

\brem
\bnum2
\item
By means of functional calculus, $f$ above is to be understood also as a mapping from
(some subspace of) the Banach algebra $\blg$ to itself.
In our particular situation, 
we simply have $f(\elblg) = \sum_k c_k \elblg^k$ 
for $f(z) = \sum_k c_k z^k$ with $c_k \in \C$ and $\norm\elblg < r$; unless otherwise noted,
the index $k$ is running over $\N$. 
Moreover, the series converges
uniformly on any $B_\rho$ with $\rho < r$; this is true for both $f$ on $\C$ and $f$ on $\blg$.
\item
Note that the lemma above is no longer true if we drop the
condition $\sfkt(0) = 0$ in the non-unital case. In fact, the simplest case
$\sfkt \ident 1$, which is perfectly analytic, gives $\sfkt \circ \varphi \ident \EINS$
for all $\varphi \in \dsa(\clg)$. But, $\EINS$ is not in $\clg \obermenge \dsa(\clg)$ 
in the non-unital case.
\item
In the unital case, one easily sees that the assumption $\sfkt(0) = 0$ 
can indeed be dropped. Setting 
$\sfktalt(z) := \sfkt(z) - \sfkt(0)$, we see that $\sfktalt(0) = 0$, hence
$\sfktalt \circ \varphi \in \dsa(\clg)$
for all $\varphi \in \dsa(\clg)$.
Now,
$\sfkt \circ \varphi \ident \sfktalt \circ \varphi + \sfkt(0) \,\EINS$ is so as well.

\enum

\erem

\bpf
Choose $c_k \in \C$ 
such that $\sfkt(z) = \sum_k c_k z^k$ for $\betrag z < r$.
From $\sfkt(0) = 0$, we get $c_0 = 0$.
As $\dsa(\clg)$ is an algebra, it contains with $\varphi$ also $\varphi^k$ 
for all $k \geq 1$. Now, $\sfkt \circ \varphi \in \quer{\dsa(\clg)}$ as
for $\supnorm\varphi < r$
\bgl
\Bigsupnorm{\sfkt \circ \varphi - \sum_{k=1}^n c_k \varphi^k}
 & \ident & \Bigsupnorm{\sum_{k=n+1}^\infty c_k \varphi^k}
 \breitrel\leq \sum_{k=n+1}^\infty \betrag{c_k} \supnorm{\varphi}^k
 \breitrel\gegen 0\,.
\egl
Consequently, as $\clg$ is Banach, we have $\sfkt \circ \varphi \in \clg$.
It remains to prove that
any of its partial derivatives is in $\clg$ again. 
For this, consider $\sfktaltalt(z) := \dot \sfkt(z) - \dot \sfkt(0)$,
which is analytic for $\betrag z < r$ and fulfills $\sfktaltalt(0) = 0$. 
Hence, $\sfktaltalt \circ \varphi$ is in $\clg$ as shown above.
As, by assumption, 
$\del_\alpha\varphi$ is in $\clg$, we have

\bgl
\del_\alpha [\sfkt \circ \varphi] 
 & = & (\dot\sfkt \circ \varphi) \cdot \del_\alpha \varphi 
 \breitrel= (\sfktaltalt \circ \varphi) \cdot \del_\alpha \varphi  + 
              \dot \sfkt(0) \, \del_\alpha \varphi
 \breitrel\in \clg.
\egl

Inductively, we see that any partial derivative of $\sfkt \circ \varphi$ is 
in $\clg$.
\qed
\epf

\section{Continuity Criterion}

\bprop
\label{prop:cont-crit-state}
Let 
\bunum
\item
$\omega$ be a state on the unital $\ast$-algebra $\alg$;
\item
$I : \dsa(\clg) + \C \, \EINS \nach \alg$ be a unital $\ast$-homomorphism.
\eunum

Then $\omega \circ I$ is continuous 
with norm
$\omega(I(\EINS))$.

\eprop
The following proof is inspired by the corresponding proof for unital $\clg$
in \cite{d117}. 

\bpf
\bunum
\item
Obviously, $\dach\omega := \omega \circ I : \dsa(\clg) + \C \, \EINS \nach \C$
is $\ast$-linear.

\item
Let $\varphi \in \dsa(\clg)$ and $r > \supnorm\varphi$. Define%
\footnote{Note that we have chosen the square-root to be holomorphic on 
$\C \setminus (-\infty,0]$ and positive on $(0,\infty)$.}
$\sfktalt(z) := \sqrt {r \pm z}$ and
$\sfkt(z) := \sfktalt(z) - \sfktalt(0)$. 
As $\sfkt$ is analytic in $0$ with convergence radius $r$ and $\sfkt(0) = 0$,
we see from Lemma \ref{lem:closed-under-analytic} that $\sfkt \circ \varphi$ is in $\dsa(\clg)$ again,
hence $\sfktalt \circ \varphi \ident \sfkt \circ \varphi + \sqrt r \,\EINS \in \dsa(\clg) + \C \, \EINS$.
\item
Let now additionally $\varphi^\ast = \varphi$.
\bunum
\item
As $\sfktalt$ is real on $(-r,r)$, 
we have $(g \circ \varphi)^\ast = g \circ \varphi$.
\item
From $(\sfktalt \circ \varphi)^\ast \cdot (\sfktalt \circ \varphi) = 
(\sfktalt \circ \varphi)^2 = \sfktalt^2 \circ \varphi = r\,\EINS \pm \varphi$, we get
\bgl
r \dach \omega(\EINS) \pm \dach\omega(\varphi)
 \breitrel= \dach \omega(r\,\EINS \pm \varphi)
 & \ident & \dach \omega\bigl((\sfktalt \circ \varphi)^\ast \cdot (\sfktalt \circ \varphi)\bigl) \\
 & = & \omega\bigl(I(\sfktalt \circ \varphi)^\ast \cdot I(\sfktalt \circ \varphi)\bigl)
 \breitrel\geq 0\,.
\egl
\item
From $\dach \omega(\EINS) = \dach \omega(\EINS^\ast \, \EINS) = 
\omega(I(\EINS)^\ast I(\EINS)) \geq 0$, we get
hence
\bgl
\betrag{\dach\omega(\varphi)} 
 & \leq & \dach \omega(\EINS) \supnorm \varphi \,,
\egl
as $r > \supnorm\varphi$ has been arbitrary. 
\eunum
\item
Let us now drop the reality assumption on
$\varphi$ and choose $\lambda \in U(1)$ with $\lambda \dach\omega(\varphi) \in \R$.
Then%
\footnote{As usual, we let 
$\re \elblg := \inv2(\elblg + \elblg^\ast)$ and $\Im \elblg := \inv{2\I} (\elblg - \elblg^\ast)$.}
\bgl
\lambda \dach\omega(\varphi)
 \breitrel= \dach\omega(\lambda \varphi)
 & = & \dach\omega(\re [\lambda \varphi]) + \I \dach\omega(\Im [\lambda\varphi])
\egl
implies $\dach\omega(\Im [\lambda\varphi]) = 0$, whence by $\supnorm{\varphi^\ast} = \supnorm{\varphi}$
\bgl
\betrag{\dach\omega(\varphi)}
 & = & \betrag{\dach\omega(\re [\lambda \varphi])} \\
 & \leq & \dach \omega(\EINS) \supnorm{\re[\lambda \varphi]}
 \breitrel\leq \dach \omega(\EINS) \supnorm{\lambda \varphi}
 \breitrel= \dach \omega(\EINS) \supnorm{\varphi}\,.
\egl
\item
The statement on the norm of $\omega \circ I$ is now obvious.
\qed
\eunum
\epf

\brem
\bunum
\item
Of course, $\omega \circ I$ restricted to $\dsa(\clg)$ is continuous as well. 
Note, however, that 
for non-unital $\clg$ the norm of this restriction 
need no longer be $\omega(I(\EINS))$; in general, it is smaller.
In fact, consider $\clg := C_0(\R)$ and $\alg := \clg + \C \, \EINS \teilmenge \boundfkt(\R)$ 
together with 
$\omega(\elclg + \lambda \, \EINS) := \lambda$. Obviously, $\omega$ is a state on $\alg$
that vanishes on $\clg$, hence on $\dsa(\clg)$ as well. Thus, if $I$ is the usual embedding,
we get $\omega \circ I \einschr{\dsa(\clg)} \ident 0$, in contrast to $\omega(I(\EINS)) = 1$.

\item
Our proof gives a sharper and more general bound for the norm than that given in the
subsection on continuity in \cite{d117}. Indeed, on the one hand, we do no longer require 
that $\omega$ is normalized nor that $\clg$ is unital; on the other hand, we were
able to drop the factor $2$ in \cite{d117}. The argument above allows also to
remove the factor $2$ in the estimate in Lemma 3.2 in \cite{lost}. There the authors 
conjectured that this might be possible, but as it had been irrelevant for their 
overall result (as it had in \cite{d117}), they refrained from proving it.

\eunum
\erem

\section{Conclusions}
The
results derived above are surely 
not the maximal extension of those claimed in \cite{d117}.
It is quite obvious, that $M$ can be replaced by a manifold or some locally convex space. 
But even more: a closer look to the proofs shows that actually
the only things we have really needed have been the 
derivation properties of the partial derivatives $\del_\alpha$ and some norm estimates.
Also, holomorphic functional calculus is much more than just replacing
$z$ in the Taylor series by some algebra element. So we expect that the findings above
are just moderate extensions of \cite{d117}.

One might now ask why we have chosen this intermediate level. 
Of course, we could have restricted ourselves just to the cases in \cite{d117}.
Or we could have searched for the maximal extension. Well, the latter is hard to 
find (if it exists at all), so we decided to let us be guided by 
further applications to be expected in loop quantum cosmology.
In particular, we are looking for information 
about models of more degrees of freedom like Bianchi I.
There, to the best of our knowledge, 
the explicit form of the restriction algebra
underlying the quantum configuration space is still completely unknown. 
It is just known that
$\clg$ is no longer a subset of $\boundfkt(\R)$, but of $\boundfkt(\R^3)$.
Nevertheless, continuity of states on $\clg$ is still given as shown above~-- 
as soon as one can prove that $\dsa(\clg)$ is dense in $\clg$.

\section*{Acknowledgements}
The author thanks the Institute for Gravitation and the Cosmos at Penn State
University for its kind hospitality. In particular, he would like to 
thank Javier Olmedo for asking the author about the
Engle-Hanusch-Thiemann paper which lead to a deeper involvement with their 
paper.


\end{document}